\documentclass[aps,prd,preprint,groupedaddress,showpacs]{revtex4}

\usepackage{epsfig}
\usepackage{graphics}
\usepackage{color}

\begin{document}

\leftmargin -2cm
\def\choosen{\atopwithdelims..}

\boldmath
\title{Prompt photon photoproduction at HERA within the framework
 of the quark Reggeization hypothesis}
\unboldmath
  \author{\firstname{V.A.} \surname{Saleev}}
\email{saleev@ssu.samara.ru}

\affiliation{ Samara State University, Ac. Pavlov St. 1, 443011
Samara, Russia}

\begin{abstract}
We study the inclusive production of isolated prompt photons within
the framework of the quasi-multi-Regge-kinematic approach, applying
the quark Reggeization hypothesis. We describe accurately and
without free parameters the transverse momentum and pseudorapidity
spectra of prompt photons in the inclusive photoproduction at the
HERA Collider. It is shown that the main mechanism of the inclusive
prompt photon production in the $\gamma p$ collisions is the fusion
of a Reggeized quark (antiquark) from the proton and a collinear
antiquark (quark) from the photon into a photon, via the effective
Reggeon-quark-gamma vertex. The fragmentation of the quark, which is
produced via the gamma-Reggeon-quark and quark-Reggeon-quark
vertices, into a photon is strongly suppressed by the isolation cone
condition and it gives a significant contribution in the region of a
large negative pseudorapidity only. At the stage of numerical
calculations we use the Kimber-Martin-Ryskin prescription for
unintegrated quark and gluon distribution functions, with the
following collinear parton densities as input: MRST for a proton and
GRV for a photon.
\end{abstract}


\maketitle \maketitle

\section{Introduction}
\label{sec:one}

It is well known that studies of production of photons with large
transverse momenta producing in the hard interaction between photon
and parton or between two partons in $\gamma p$ collisions,
so-called prompt photon production, provide precision tests of
perturbative quantum chromodynamics (QCD) as well as information on
the parton densities within a proton and a photon.

Also, these studies are our potential for the observation of a new
dynamical regime, namely the high-energy Regge limit, which is
characterized by the following condition
$\sqrt{S}>>\mu>>\Lambda_{QCD}$, where $\sqrt{S}$ is the total
collision energy in the center of mass reference frame,
$\Lambda_{QCD}$ is the asymptotic scale parameter of QCD, $\mu$ is
the typical energy scale of the hard interaction. At this
high-energy limit, the contribution from the partonic subprocesses
involving $t-$channel parton (quark or gluon) exchanges to the
production cross section can become dominant. In the region under
consideration, the transverse momenta of the incoming partons and
their off-shell properties can no longer be neglected, and we deal
with "Reggeized" $t-$channel partons.

We consider the quasi-multi-Regge kinematics (QMRK) approach
\cite{FadinLipatov96}, which is similar to the $k_T-$factorization
approach \cite{KTALL} conceptually, as the theoretical framework for
this kind of high-energy phenomenology. Our previous analysis of
charmonium and bottomonium production at the Fermilab Tevatron and
DESY HERA Colliders \cite{KniehlSaleevVasin}, and the study of
prompt photon production at the Fermilab Tevatron \cite{Saleev2008},
using the high-energy factorization scheme have shown the
appreciation of the method even at the leading order (LO) in the
strong-coupling constant $\alpha_s$ in compare with the experimental
data.  We have found that the essential features produced by the
high-energy factorization scheme at the LO are being at the
next-to-leading (NLO) in the conventional collinear parton model.

Contrary to the $k_T-$factorization scheme, based on the well-known
prescription for off-shell $t-$channel gluons \cite{KTALL}, the QMRK
approach seems to be more proper theoretically \cite{Lipatov95}. In
this approach we work with gauge invariant amplitudes and use the
factorization hypothesis, which is proved in the
leading-logarithmic-approximation (LLA). In the QMRK approach we can
do calculationSs with Reggeized quarks \cite{Saleev2008}, that
presents an open question in the $k_T-$factorization approach,
following which we can operate correctly with off-shell gluons only.
Recently it was shown also, that the calculation in the NLO in
strong coupling constant within the framework of QMRK approach can
be done \cite{NLO}.

In this paper we study the inclusive photoproduction of prompt
photon within the framework of the QMRK approach, applying the quark
and gluon  Reggeization hypothesis
\cite{FadinLipatov96,FadinSherman}. We describe accurately and
without free parameters the transverse momentum and pseudorapidity
spectra of prompt photons in the inclusive photoproduction at the
HERA Collider \cite{H1gamma,ZEUSgamma}. It is shown that the main
mechanism of the inclusive prompt photon production in the $\gamma
p$ collisions is the fusion of a Reggeized quark (antiquark) from
the proton and a collinear antiquark (quark) from the photon into a
photon, via the effective Reggeon-quark-gamma vertex. The
fragmentation of the quark, which is produced via the
gamma-Reggeon-quark and quark-Reggeon-quark vertices, into a photon
is strongly suppressed by the isolation cone condition and it gives
a significant contribution in the region of a large negative
pseudorapidity only. At the stage of numerical calculations we use
the Kimber-Martin-Ryskin prescription for unintegrated quark and
gluon distribution functions \cite{KMR}, with the following
collinear densities as input: MRST for a proton \cite{MRST} and GRV
for a photon \cite{GRVph}.

 This paper is organized as follows.
In Sec.~\ref{sec:two}, the relevant Reggeon-Reggeon-particle  and
particle-Reggeon-particle  effective vertices are presented and
discussed. In Sec.~\ref{sec:three}, we describe the inclusive photon
photoproduction at the HERA Collider. In Sec.~\ref{sec:foure}, we
compare our approach with the similar calculations in the
$k_T-$factorization scheme. In Sec.~\ref{sec:five}, we summarize our
conclusions.

\section{Basic Formalism}
\label{sec:two}

In the phenomenology of strong interactions at high energies, it is
necessary to describe the QCD evolution of the parton distribution
functions of the colliding particles starting with some scale
$\mu_0$, which controls a nonperturbative regime, to the typical
scale $\mu$ of the hard-scattering processes, which is typically of
the order of the transverse mass $M_T=\sqrt{M^2+|{\bf p}_T|^2}$ of
the produced particle with mass $M$ and transverse momentum ${\bf
p}_T$. In the region of very high energies, in the so-called Regge
limit, the typical ratio $x=\mu/\sqrt{S}$ becomes very small,
$x\ll1$. This leads to large logarithmic contributions of the type
$[\alpha_s\ln(1/x)]^n$ in the resummation procedure, which is
described by the BFKL evolution equation \cite{BFKL} or other
BFKL-like ones for an unintegrated gluon (quark) distribution
functions $\Phi_{g,q}(x,|{\bf q}_T|^2,\mu^2)$. Correspondingly, in
the QMRK approach \cite{FadinLipatov96}, the initial-state
$t$-channel gluons and quarks are considered as Reggeons, or as
Reggeized gluons $(R)$ and as Reggeized quarks $(Q)$. They are
off-mass shell and carry finite transverse two-momenta ${\bf q}_T$
with respect to the hadron beam from which they stem.

Recently, in Ref.~\cite{KTAntonov}, the Feynman rules for the
effective theory based on the non-Abelian gauge-invariant action
\cite{Lipatov95} were derived for the induced and some important
effective vertices. However, these rules include processes with
Reggeized gluons in the initial state only. In case of $t$-channel
quark exchange processes such rules are absent today, in spite of
that the gauge-invariant action for Reggeized quarks is known
\cite{LipatoVyazovsky}, and it is needed to construct effective
vertices with the Reggeized quarks, using QMRK approach
prescriptions, every time from the beginning. Of course, a certain
set of Reggeon-Reggeon-particle effective vertices is known, for
example, for the transitions $R R\to g$ \cite{LipatovFadin89},
$Q\bar Q\to g$ \cite{FadinSherman} and $\gamma^*Q\to q$
\cite{Saleev2008}.

Roughly speaking, the Reggeization of amplitudes is a trick that
offers an opportunity to take into account efficiently  large
radiation corrections to the processes under Regge limit condition
outside the collinear approximation. The particle Reggeization is
known in high energy quantum electrodynamics (QED) for electrons
only \cite{GellMann} and for gluon and quarks in QCD
\cite{BFKL,FadinSherman}.

The effective vertex $C_{\bar Q Q}^g(q_1,q_2)$, which describes the
production of a single gluon with momentum $k(g)=q_1(Q)+q_2(\bar Q)$
in the Reggeized quark and Reggeized antiquark  annihilation was
obtained in Ref.~\cite{FadinSherman,FadinBogdan} and it can be
presented, omitting the color and Lorentz indexes in the left side,
as follows:
\begin{equation}
C_{\bar Q Q}^g(q_1,q_2)=D_a^\mu(q_1,q_2)=-g_s T^a \left( \gamma^\mu
-\frac{2P_1^\mu }{x_2S}\hat{q_2}-\frac{2P_2^\mu
}{x_1S}\hat{q_1}\right)\label{eq:QQg},
\end{equation}
where $P_{1,2}$ are colliding hadron momenta,
$P_{(1,2)}=E_{(1,2)}(1,\vec 0,\pm 1)$ and $S=2(P_1P_2)=4E_1E_2$. The
Reggeized quark momenta are $q_1=x_1P_1+q_{1T}$,
$q_2=x_2P_2+q_{2T}$, and $q_{(1,2)T}=(0,\vec q_{(1,2)T},0)$. It is
obvious that this vertex satisfies the gauge-invariant condition
$D_{a}^\mu(q_1,q_2) k_\mu=0$.

Like the gluon production, the effective vertex for photon
production can be written in the form:
\begin{equation}
C_{\bar Q Q}^\gamma(q_1,q_2)=F^\mu(q_1,q_2)=-e e_q \left( \gamma^\mu
-\frac{2P_1^\mu }{x_2S}\hat{q_2}-\frac{2P_2^\mu
}{x_1S}\hat{q_1}\right)\label{eq:QQgamma},
\end{equation}
where $e=\sqrt{4\pi\alpha}$ is the electromagnetic coupling
constant, $e_q$ is the quark electric charge.


The effective vertex, which describes the production of a quark with
momentum $k=q_1(\gamma^\star)+q_2(Q)$ in a virtual photon collision
with a Reggeized quark can be presented as follows
\cite{Saleev2008}:
\begin{equation}
C_{\gamma Q}^q(q_1,q_2)=G^\mu(q_1,q_2)=-e e_q\left(
\frac{q_1^2}{q_1^2+q_2^2}\gamma^\mu-\frac{2k^\mu}{q_1^2+q_2^2}\hat{q_2}
+\frac{2q_2^2 x_2
P_2^\mu}{(q_1^2+q_2^2)^2}\hat{q_2}\right)\label{eq:gammaQq},
\end{equation}
where $q_2=x_2P_2+q_{2T}$ is the Reggeized quark four-momentum and
the gauge invariance condition is given by $G^\mu(q_1,q_2)q_{1\mu}
=0.$ For a mass-shell photon the vertex (\ref{eq:gammaQq}) can be
presented in the form:
\begin{equation}
G^\mu(q_1,q_2)=2 e e_q\frac{\hat q_{2T}q_{2T}^\mu}{q_{2}^2}.
\end{equation}

The effective Reggeon-Reggeon-particle vertex, which describes the
production of a quark with momentum $k(q)=q_1(Q)+q_2(R)$ in the
Reggeized quark  and Reggeized gluon collision can be presented as
follows:
\begin{equation}
C_{R Q}^q(q_1,q_2)=V(q_1,q_2)=-g_s T^a \hat \varepsilon_T
(q_2)\label{eq:RQq},
\end{equation}
where $\hat \varepsilon_T(q_2)=\gamma_\mu \varepsilon^\mu(q_2)$ and
$\varepsilon^\mu(q_2)=\displaystyle{\frac{q_{2T}^\mu}{|\vec q_{2T}
|}}$.

Let us note that contrary to the definition used in
Ref.~\cite{FadinBogdan}, we do not include quark spinors in our
equations for the effective vertices (\ref{eq:QQg})-(\ref{eq:RQq}).

The relevant squared matrix elements can be presented as follows:
\begin{eqnarray}
\overline{|M(Q\bar Q\to \gamma )|^2}&=&\frac{4}{3}\pi \alpha e_q^2
(t_1+t_2),
\\
\overline{|M(R Q\to q)|^2}&=&\frac{2}{3}\pi \alpha_s (t_1+t_2+ 2
\sqrt{t_1 t_2} \cos\phi ),
\\
\overline{|M(\gamma Q\to q)|^2}&=&8\pi \alpha e_q^2 t_2,
\\
\overline{|M(R R\to g)|^2}&=&\frac{3}{2}\pi \alpha_s (t_1+t_2+ 2
\sqrt{t_1 t_2} \cos\phi ),
\end{eqnarray}
where $t_1={{\vec q}_{1T}}^2$, $t_2={{\vec q}_{2T}}^2$, and $\phi$
is the azimuthal angle between $\vec q_{1T}$ and $\vec q_{2T}$.

 \boldmath
\section{Inclusive prompt photon photoproduction at HERA}
\unboldmath \label{sec:three}

In this part we consider the inclusive prompt photon production at
the HERA Collider in the region of a small exchange photon
virtuality $Q^2<1$ GeV$^2$, i.e. in the photoproduction processes.
Working in the LO QMRK approach we describe data applying the quark
and gluon Reggeization hypothesis. There are two main mechanisms of
prompt photon production in this approach: the production of direct
photons and the production of photons via fragmentation. In the
first case, photons are produced directly via fusion of a Reggeized
quark and a Reggeized antiquark, which stem from a proton (photon)
and a photon (proton):
\begin{equation}Q_{\gamma}(\bar Q_{\gamma}) + \bar Q_{p} (Q_p)\to \gamma.\label{eq:direct}
\end{equation}
In the second case, photons are produced in the quark or gluon
fragmentation into the  photon, which is described by the parton to
photon fragmentation functions $D_{q\to \gamma}(z,\mu^2)$ and
$D_{g\to \gamma}(z,\mu^2)$ \cite{DukeOwens}. In the discussed here
task the gluon fragmentation into the photon is unimportant and we
will take into account the quark(antiquark) fragmentation only:
\begin{eqnarray} &&\gamma + Q_p(\bar Q_p) \to q (\bar q)\to \gamma,\label{eq:resolved1}\\
&&Q_\gamma (\bar Q_\gamma)+ R_p\to q (\bar q)\to \gamma,\label{eq:resolved2}\\
 &&R_\gamma + Q_p (\bar Q_p)\to q (\bar q)\to
\gamma. \label{eq:resolved3}
\end{eqnarray}

The factorization formulas for the production cross section of the
direct photons (\ref{eq:direct}) can be written as follows:
\begin{eqnarray}
d\sigma^{dir}(eN\to \gamma X)&=&\sum_{q,\bar q}\int dy \int dx_1\int
\frac{d^2q_{1T}}{\pi} \int dx_2\int \frac{d^2q_{2T}}{\pi}
G_{\gamma/e}(y)\Phi_{q/\gamma}(x_1,t_1,\mu^2) \times\label{f:direct}\\
\nonumber &&\times \Phi_{\bar q/p}(x_2,t_2,\mu^2)d\hat\sigma (Q \bar
Q\to \gamma).
\end{eqnarray}

In the case of fragmentation production mechanism we have two
formulas for the "direct" (\ref{eq:resolved1}) and the "resolved"
(\ref{eq:resolved2},\ref{eq:resolved3}) contributions:
\begin{eqnarray}
d\sigma^{fr,dir}(eN\to \gamma X)&=&\sum_{q,\bar q}\int dz\int dy
\int dx_2\int \frac{d^2q_{2T}}{\pi} G_{\gamma/e}(y)\times\\
\nonumber &&\times \Phi_{g/p}(x_2,t_2,\mu^2)d\hat\sigma (\gamma Q
\to q)D_{q\to\gamma}(z,\mu^2)\label{f:direct2},
\end{eqnarray}
\begin{eqnarray}
d\sigma^{fr,res}(eN\to \gamma X)&=&\sum_{q,\bar q}\int dz\int dy
\int dx_1\int \frac{d^2q_{1T}}{\pi} \int dx_2\int
\frac{d^2q_{2T}}{\pi} G_{\gamma/e}(y)\times\\ \nonumber &&\times
\Phi_{q/\gamma}(x_1,t_1,\mu^2) \Phi_{g/p}(x_2,t_2,\mu^2)d\hat\sigma
(Q R\to q)D_{q\to\gamma}(z,\mu^2)\label{f:resolved}.
\end{eqnarray}
Here $\Phi_{q,g/\gamma}(x_1,t_1,\mu^2)$ is the unintegrated quark or
gluon distribution in a photon, which is calculated using
KMR\cite{KMR} prescription, the same as the unintegrated quark or
gluon distribution in a proton
$\Phi_{q,g/p}(x_2,t_2,\mu^2)$,\textcolor{blue}{;} $G_{\gamma/e}(y)$
denotes the photon flux integrated over $Q^2$ from the kinematic
limit of $Q^2_{min}=m_e^2 y^2/(1-y)$ to the upper limit $Q^2_{max} =
1$ GeV$^2$; $y=W^2/S_{eN}$, $W^2=2(P_\gamma P_N)$,
$S_{eN}=2(P_eP_N)=4 E_N E_e$, $E_N$ and $E_e$ are the proton and
electron energies in the laboratory frame. We use the following
collinear densities  as input: MRST \cite{MRST} for a proton and GRV
\cite{GRVph} for a photon. The exact formula for $G_{\gamma/e}(y)$
is taken as follows:
\begin{equation}
G_{\gamma/e}(y)=\frac{\alpha}{2
\pi}\biggl[\frac{1+(1-y)^2}{y}\log{\frac{Q^2_{max}}{Q^2_{min}}}+2
m_e^2 y
\bigl(\frac{1}{Q^2_{min}}-\frac{1}{Q^2_{max}}\bigr)\biggr]\mbox{.}
\end{equation}
%
To calculate transverse momentum and pseudorapidity spectra in the
photoproduction of the direct photon we use the following formula,
which is obtained from (\ref{f:direct}):
\begin{eqnarray}
\frac{d\sigma}{dp_Td\eta}(eN\to \gamma
X)&=&\frac{1}{p_T^3}\sum_{q,\bar q}\int dy \int d\phi_1\int dt_1
G_{\gamma/e}(y)\Phi_{\bar q,q/\gamma}(x_1,t_1,\mu^2)\times\\
\nonumber && \times \Phi_{q,\bar q/p}(x_2,t_2,\mu^2)\overline{|M(Q
\bar Q\to \gamma)|^2},
\end{eqnarray}
where $p=(p_0,{\bf p}_T,p_z)$ is the produced photon four-momentum,
$P_e=E_e(1,{\bf 0},-1)$, $P_N=E_N(1,{\bf 0},1)$,
$$x_1=\frac{p_0-p_z}{2E_ey},\quad x_2=\frac{p_0+p_z}{2E_N},\quad t_2=
t_1-2{p_T}\sqrt{t_1}\cos (\phi_1)+{p_T^2}.$$

In case of the fragmentation production mechanism the master
formulas can be obtained from (\ref{f:direct2}), and
(\ref{f:resolved}) and they have the following forms:
\begin{eqnarray}
\frac{d\sigma}{dp_Td\eta}(eN\to \gamma
X)&=&\frac{2\pi}{p_T^3}\sum_{q,\bar q}\int dy G_{\gamma/e}(y)\Phi_{q,\bar q/p}(x_2,t_2,\mu^2)\times\nonumber\\
&\times& z^3 D_{q\to \gamma}(z,\mu^2)\overline{|M(\gamma Q\to q|^2},
\end{eqnarray}
where
$$z=\frac{p_0-p_z}{2yE_e}, \quad x_2=\frac{p_0+p_z}{2zE_N}, \quad t_2=\frac{p_T^2}{z^2}.$$

\begin{eqnarray}
\frac{d\sigma}{dp_Td\eta}(eN\to \gamma
X)&=&\frac{1}{p_T^3}\sum_{q,\bar q}\int dy \int d\phi_1\int dt_1\int
dz~ G_{\gamma/e}(y)\Phi_{q,\bar q}^\gamma(x_1,t_1,\mu^2)\Phi_{g}^{p}(x_2,t_2,\mu^2)\times\nonumber\\
&\times& z^2 D_{q\to \gamma}(z,\mu^2)\overline{|M(QR\to q|^2},
\end{eqnarray}
where
$$x_1=\frac{p_0-p_z}{2yzE_e}, \quad x_2=\frac{p_0+p_z}{2zE_N}, \quad t_2=
t_1-2\frac{p_T}{z}\sqrt{t_1}\cos (\phi_1)+\frac{p_T^2}{z^2}.$$

The requirement of particle Reggeization demands fulfilment of the
conditions $x_{1,2}\ll 1$ for partons both from a proton and a
photon in the discussed here processes. As example, in case of the
direct photon production we can write:
$$x_1=\frac{p_T e^{-\eta}}{2 y E_e} \qquad\mbox{and}\qquad
x_2=\frac{p_T e^\eta}{2E_N},$$ where for the ZEUS \cite{ZEUSgamma}
and H1 \cite{H1gamma} measurements: $E_e=27.5$ GeV, $E_N=820 (920)$
GeV, $|\eta|\leq 1$, $5< p_T< 10$ GeV and $0.2<y<0.7 (0.9)$. If we
take the fixed median values: $y=0.5$, $\eta=0$, $p_T=7.5$ GeV and
$E_N=920$ GeV, we obtain that $x_1\simeq 0.25$ and $x_2\simeq
0.004$. In such a way, the particle Reggeization should be used for
partons from a proton only, and partons from a photon have been
considered in the collinear approximation, as usually in the
conventional parton model.

In that case all master formulas can be rewrited using the following
prescriptions: \begin{eqnarray} &&\int dt_1
\Phi_{q/\gamma}(x_1,t_1,\mu^2) \Rightarrow x_1
F_{q/\gamma}(x_1,\mu^2), \nonumber\\ &&\int d\phi_1 \Rightarrow
2\pi,\nonumber \end{eqnarray} where $F_{q/\gamma}(x_1,\mu^2)$ is the
 collinear parton density in a photon \cite{GRVph}, and we need to take $t_1=0$ in the all
kinematic relations and in the squared matrix elements.

 To compare the theoretical predictions and the data we need to take into
 account carefully kinematic conditions and cuts of the discussed experiments
 \cite{H1gamma,ZEUSgamma}. The special attention should be attracted
 to the isolation cone condition $E_T^{cone}<0.1~ E_T$
 \cite{H1gamma}, where $E_T$ is the total transverse energy
 of a jet with the detected photon, $E_T^{cone}$ is the non-photon
 energy in this jet. It is obvious that the isolation cone
 condition needs the inequality $0.9<z<1$ to be fulfilled in the
case of fragmentation
 production mechanism. This cut on the fragmentation parameter $z$ strongly
 suppresses the contribution of the fragmentation production.

However, the isolation cone condition should be taken into account
and in the production of direct photons too. The measured transverse
energy $E_T$ includes additional non-photon part which is order of
$E_T^{cone}$ \cite{H1gamma}, thus we can put
$p_T=E_T-E_T^{cone}\simeq 0.9~ E_T$, if the theoretical prediction
corresponds to the photon transverse momentum $p_T$.

\begin{figure}[ht]
\begin{center}
\includegraphics[width=.9\textwidth, clip=]{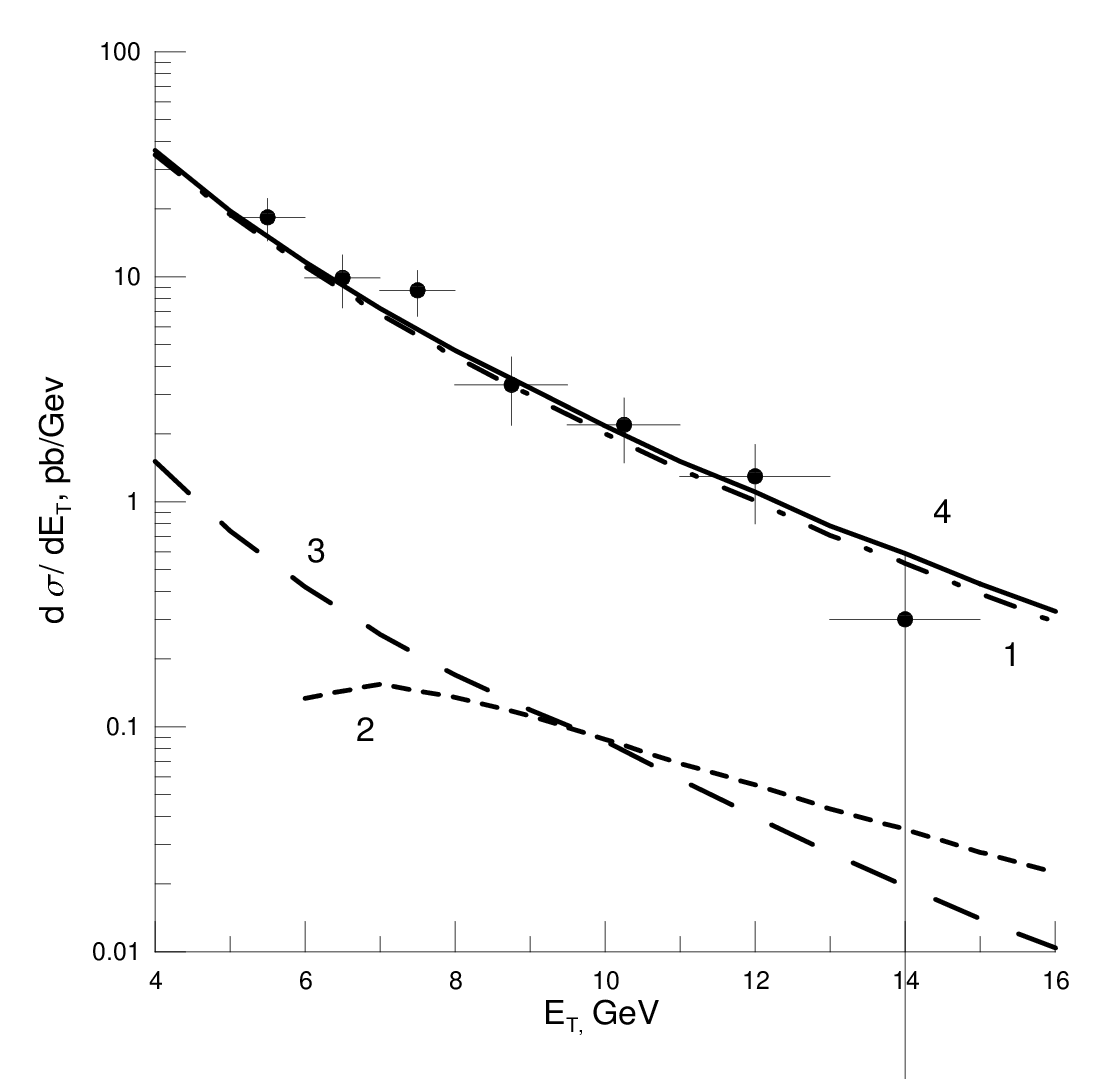}
\end{center}
\caption{The $E_T-$spectrum of prompt photon at $\sqrt{S}=300$ GeV,
$-0.7<\eta<0.9$, and $0.2<y<0.9$. The curve 1 is the resolved-direct
production, 2 is the direct-fragmentation production, 3 is the
resolved-fragmentation production, and 4 is their sum. The data are
from ZEUS Collaboration \cite{ZEUSgamma}.\label{fig:1}}
\end{figure}

\begin{figure}[ht]
\begin{center}
\includegraphics[width=.9\textwidth, clip=]{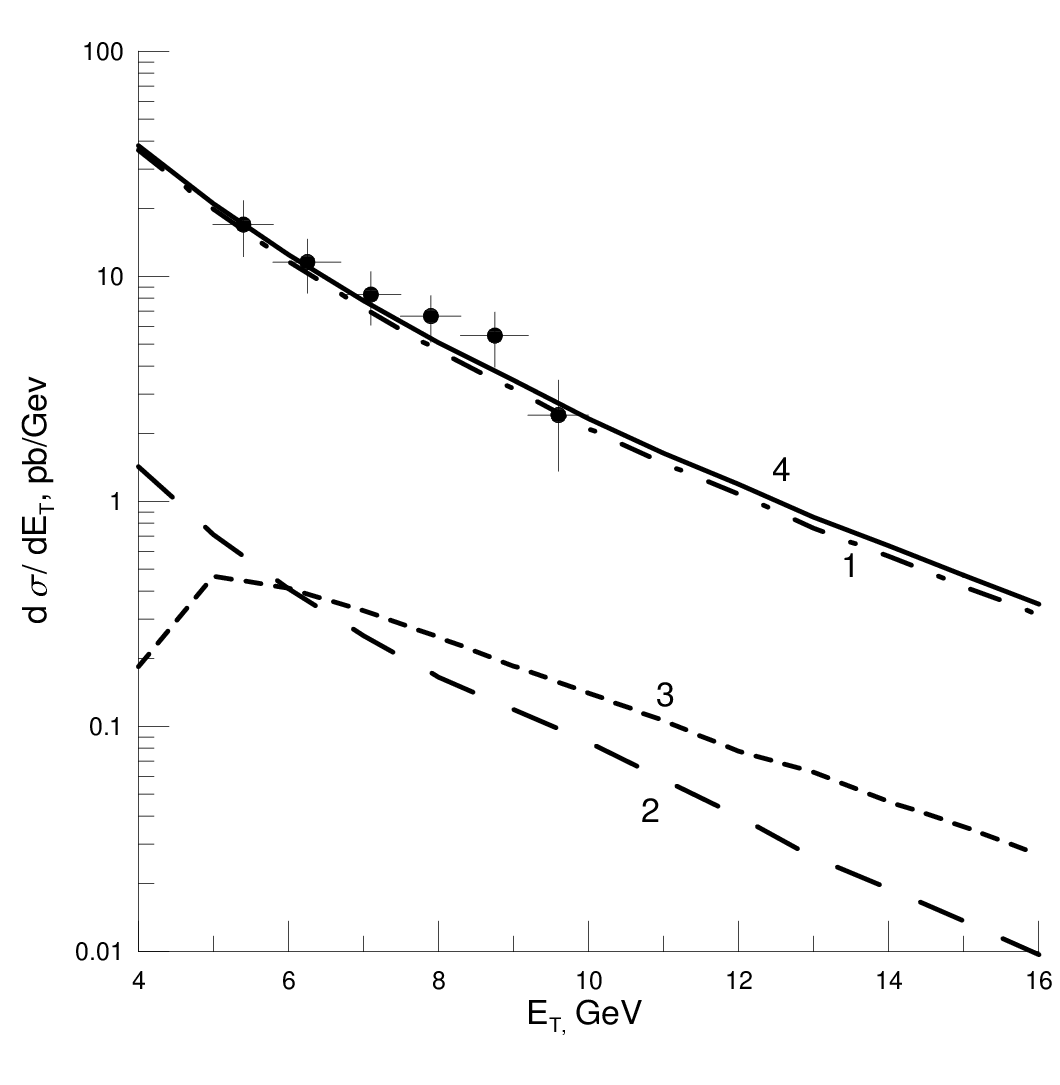}
\end{center}
\caption{The $E_T-$spectrum of prompt photon at $\sqrt{S}=319$ GeV,
$-1.0<\eta<0.9$, and $0.2<y<0.7$. The curve 1 is the resolved-direct
production, 2 is the direct-fragmentation production, 3 is the
resolved-fragmentation production, and 4 is their sum. The data are
from H1 Collaboration \cite{H1gamma}. \label{fig:2}}
\end{figure}

\begin{figure}[ht]
\begin{center}
\includegraphics[width=.9\textwidth, clip=]{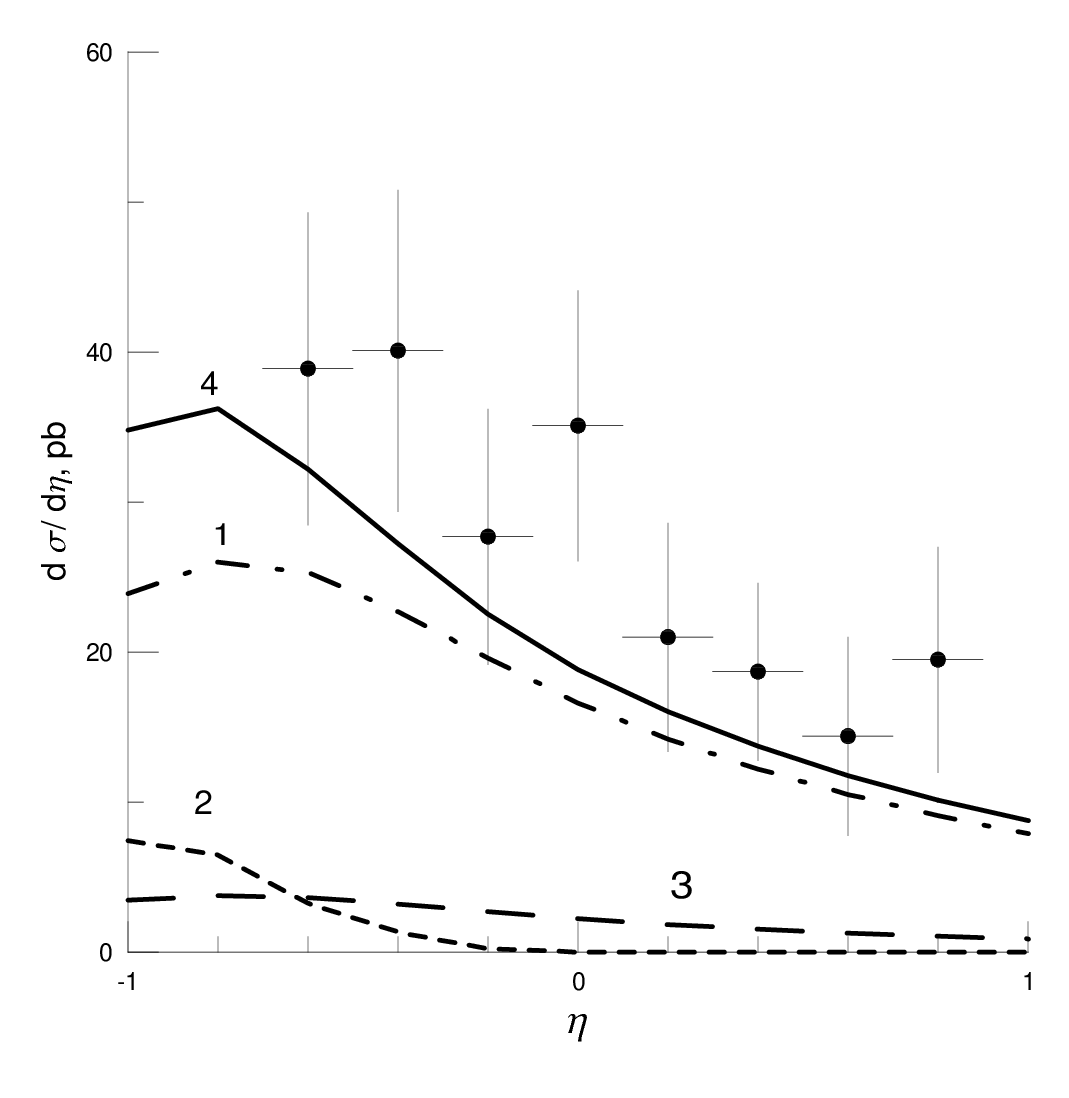}
\end{center}
\caption{The $\eta-$spectrum of prompt photon at $\sqrt{S}=300$ GeV,
$5<E_T<10$ GeV, and $0.2<y<0.9$. The curve 1 is the resolved-direct
production, 2 is the direct-fragmentation production, 3 is the
resolved-fragmentation production, and 4 is their sum. The data are
from ZEUS Collaboration \cite{ZEUSgamma}.\label{fig:3}}
\end{figure}

\begin{figure}[ht]
\begin{center}
\includegraphics[width=.9\textwidth, clip=]{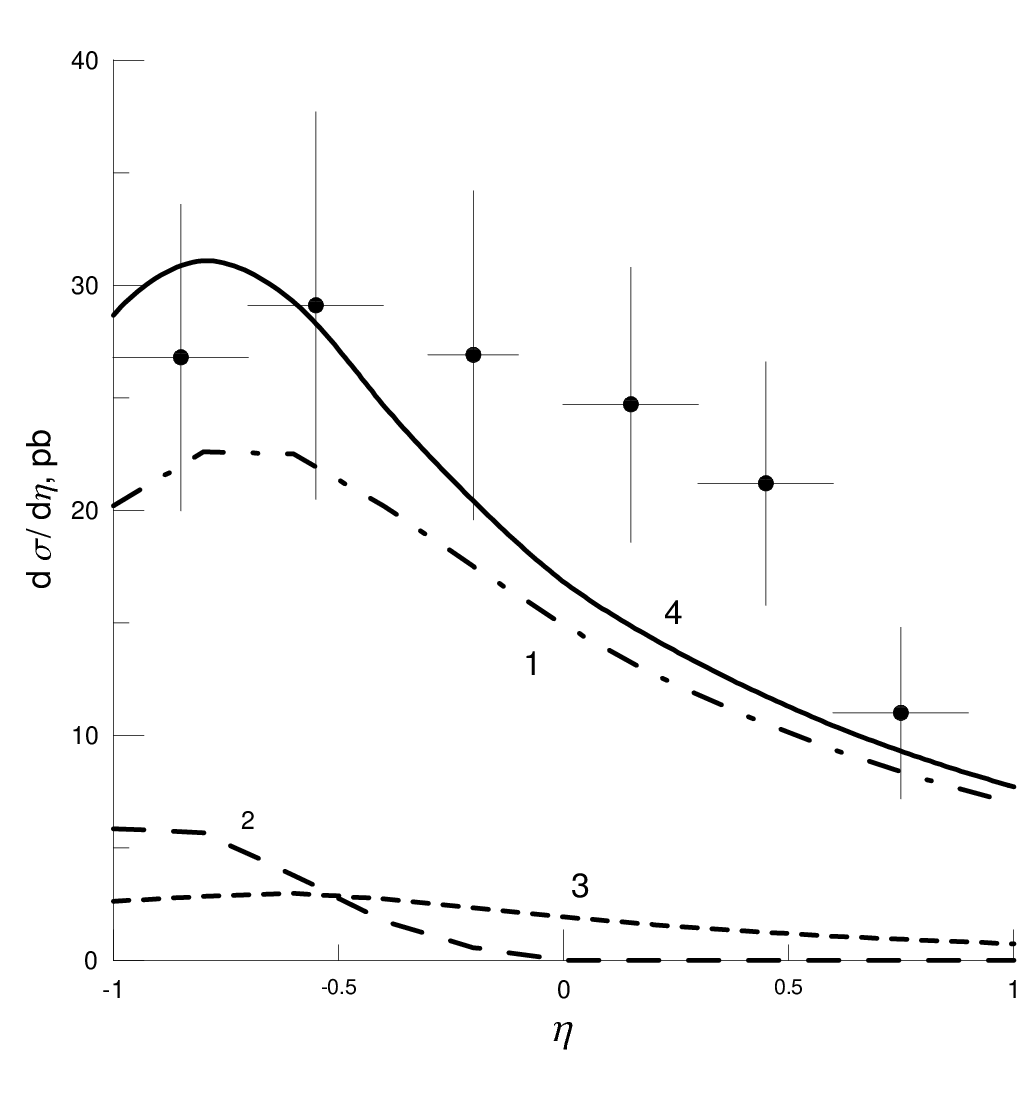}
\end{center}
\caption{The $\eta-$spectrum of prompt photon at $\sqrt{S}=319$ GeV,
$5<E_T<10$ GeV, and $0.2<y<0.7$. The curve 1 is the resolved-direct
production, 2 is the direct-fragmentation production, 3 is the
resolved-fragmentation production, and 4 is their sum. The data are
from H1 Collaboration \cite{H1gamma}. \label{fig:4}}
\end{figure}

 We compare the results of our calculations with the
data from  ZEUS \cite{ZEUSgamma} and  H1 \cite{H1gamma}
Collaborations. Figures (\ref{fig:1})--(\ref{fig:4}) demonstrate an
agreement between the theory and the data, especially for the
transverse-energy spectra. We use the following notation for the
curves: 1 is the resolved production of the direct photons in the
subprocesses (\ref{eq:direct}); 2 is the direct production via the
fragmentation in the subprocesses (\ref{eq:resolved1}); 3 is the
resolved production via the fragmentation in the subprocesses
(\ref{eq:resolved2}) and (\ref{eq:resolved3}); 4 is the sum of all
contributions.

In the transverse energy distribution the mechanism of the direct
photon production (curve 1) is dominant at the all $E_T^\gamma$ and
the fragmentation contribution (curves 2 and 3) is only about $5 \%$
of the first one, that is in accordance to the previous calculations
in the NLO of the collinear parton model \cite{NLOgamma}.

The fragmentation production mechanism gives significant
contribution only at the large negative values of $\eta$ in the
pseudorapidity distribution. We see that in this region the
$\eta-$spectrum can be described correctly if we take into account
both the direct (curve 1) and fragmentation (curves 2 and 3)
mechanisms. We have obtained that the data lie slightly higher our
predictions in case of the ZEUS measurements at all values of
pseudorapidity, see Fig.(\ref{fig:3}). In case of the H1
measurements the small disagreement is observed in the central
region of pseudorapidity $(\eta\approx 0)$ only, see
Fig.(\ref{fig:4}).

The small visible disagreement for the $\eta-$spectra should be
explained that we do not use in our calculation the cone isolation
condition in the pseudorapidity and azimuthal angle plane
\begin{equation}
(\eta_j-\eta)^2+(\phi_j-\phi)^2\leq R,\label{eq:cone}
\end{equation}
where $R=1$ \cite{ZEUSgamma,H1gamma}, $\eta_j$ and $\phi_j$ are the
pseudorapidity and azimuthal angle of a hadron from the detected
photon jet. Contrary to the collinear parton model calculation,
where the LO processes are $2\to 2$ ($\gamma q\to \gamma q$, $gq\to
\gamma q$, ...), the LO processes in the QMRK approach are $2\to 1$
($Q\bar Q\to \gamma$, ...). The all quark and gluon jets are
absorbed in the unintegrated distribution functions of a proton and
a photon. Thus, our prediction corresponds correctly the production
of  "pure" direct photons and it is unclear how to take into account
the cone isolation condition (\ref{eq:cone}). Comparing the
theoretical predictions and the data we should take into account
that the ZEUS and H1 data contain arbitrary small part ($\sim 10
\%$) of the non-photon events.

\boldmath
\section{Discussion}
\unboldmath \label{sec:foure}

In this section, we compare our results with the previous study in
the $k_T-$factorization scheme, which is similar to the QMRK
approach.

The calculation of the inclusive prompt photon production at the
HERA Collider was performed using off-shell amplitudes involving
initial quarks in the Ref.~\cite{ZotovLipatovGamma}. In  that paper,
a Compton process of the off-shell quark -- photon scattering
$q^\star + \gamma \to q + \gamma $ was considered as the LO direct
photon production process. We especially used here different
denotations for the Reggeized particles ($Q,R$) and off-shell
particles ($q^\star, g^\star$). It means that in
Ref.~\cite{ZotovLipatovGamma}, the conventional QCD vertices are
used and the authors obtained gauge uninvariant amplitudes. The
gauge invariant condition for amplitudes controls the correct $k_T$
dependence of the relevant cross sections. The arbitrary inclusion
of the $k_T$ effects in the kinematics only, without the using of
the correct vertices for Reggeized particle interactions can be
considered as a phenomenological trick, but such inclusion does not
have predictive power. The additional important remark to the
approach used in Ref. \cite{ZotovLipatovGamma}: the contribution of
the LO process $q^\star \bar q^\star\to \gamma$ ($Q\bar Q\to\gamma$)
was ignored. We see that this contribution is dominant and the
comparison of theoretical predictions with data without taking it
into account is ungrounded.

The Compton scattering of photons on Reggeized quarks,
\begin{equation}
\gamma + Q \to \gamma + q,
\end{equation}
is a NLO process in the QMRK approach.  To calculate it we need to
solve the same problems as for other NLO Reggeon-Reggeon to
particle-particle processes in the QMRK approch. Firstly, it is
necessary to obtain the relevant effective vertex. This task has not
been solved yet. Of course, the relevant process is LO if we study
the associated photon plus jet, both with large transverse momenta
production. However, in that case the effective vertex should be
constructed too.

 \boldmath
\section{Conclusion}
\unboldmath \label{sec:five}

It is shown that it is possible to describe data in the LO QMRK
approach  for prompt photon spectra for the inclusive
photoproduction  in the  high-energy $\gamma p$ collisions at the
HERA Collider. The used here scheme of a calculation have been
suggested recently in our paper \cite{Saleev2008}, in which we
described successfully the deep inelastic structure functions $F_2$
and $F_L$ of the lepton-proton scattering and prompt photon spectra
for the inclusive production in the $p \bar p$ collisions at the
Tevatron Collider. Our results demonstrate that the quark
Reggeization hypothesis is a very powerful tool in the high energy
phenomenology for the hard processes involving quark exchanges.

\section{Acknowledgements}
We thank  L.~Lipatov and M.~Ryskin for the critical remarks and
useful discussion of the questions under consideration in this
paper. We also thank A.~Shipilova for help in numerical calculation
of unintegrated structure functions in a photon, using KMR
prescription.

\newpage

\end{document}